\documentclass[runningheads]{llncs}
\usepackage{booktabs}
\usepackage{hyperref}
\usepackage{float} 
\usepackage[T1]{fontenc}
\usepackage{listings}
\usepackage{xcolor}
\usepackage{caption}
\definecolor{codegreen}{rgb}{0,0.6,0}
\definecolor{codegray}{rgb}{0.5,0.5,0.5}
\definecolor{codepurple}{rgb}{0.58,0,0.82}
\definecolor{backcolour}{rgb}{0.95,0.95,0.92}
\lstdefinestyle{mystyle}{
    backgroundcolor=\color{backcolour},   
    commentstyle=\color{codegreen},
    keywordstyle=\color{magenta},
    numberstyle=\tiny\color{codegray},
    stringstyle=\color{codepurple},
    basicstyle=\ttfamily\footnotesize,
    breakatwhitespace=false,         
    breaklines=true,                 
    captionpos=b,                    
    keepspaces=true,                 
    numbers=left,                    
    numbersep=5pt,                  
    showspaces=false,                
    showstringspaces=false,
    showtabs=false,                  
    tabsize=2
}

\usepackage{graphicx}
\usepackage{subcaption}

\usepackage{color}

\begin{document}
\title{Assessing Tenstorrent's RISC-V MatMul Acceleration Capabilities}

% \author{Anonymous CADL submission}

\author{Hiari {Pizzini Cavagna}\inst{1}\orcidID{0009-0005-2768-0418} \and
Daniele Cesarini \inst{2}\orcidID{0000-0003-1294-372X}\and
Andrea Bartolini \inst{1}\orcidID{0000-0002-1148-2450}}

\authorrunning{H. Pizzini Cavagna, D. Cesarini, A. Bartolini}
\institute{
University of Bologna, Bologna BO 40136, Italy\\
\email{\{hiari.pizzinicavagna, a.bartolini\}@unibo.it}\\ 
\and
Cineca, Casalecchio di Reno BO 40033, Italy\\
\email{d.cesarini@cineca.it}\\}

\maketitle              % typeset the header of the contribution

\begin{abstract}
The increasing demand for generative AI as Large Language Models (LLMs) services has driven the need for specialized hardware architectures that optimize computational efficiency and energy consumption. This paper evaluates the performance of the Tenstorrent Grayskull e75 RISC-V accelerator for basic linear algebra kernels at reduced numerical precision, a fundamental operation in LLM computations. We present a detailed characterization of Grayskull's execution model, %highlighting significant performance differences between the initial and subsequent executions due to compilation and data movement overheads. Our analysis explores how factors such as processor 
grid size, matrix dimensions, data formats, and numerical precision impact on computational efficiency. Furthermore, we compare Grayskull's performance against state-of-the-art architectures with tensor acceleration, including Intel Sapphire Rapids processors and two NVIDIA GPUs (V100 and A100). Whilst NVIDIA GPUs dominate raw performance, Grayskull demonstrates a competitive trade-off between power consumption and computational throughput, reaching a peak of 1.55 TFLOPs/Watt with BF16.

\end{abstract}
\section{Introduction}
Large Language Models, based on Transformer architecture \cite{vaswani2023attention}, achieve state-of-the-art (SoA) results in various NLP tasks. Their ability to generalize a multitude of tasks is related with their size, allowing them to achieve excellent results even for tasks for which they have not been directly trained. Alongside with the growing demand for LLMs services, there has been an increasing need for architectures that allow for efficient use in terms of both performance and energy consumption. With billions—or even hundreds of billions—of parameters, models like GPT-4 demand substantial memory and computational resources, making efficient deployment a significant challenge.

% In literature several approaches to parallelize and handling the data transfers in ML kernels results in different data-stationarieties (weights, activations, etc). Thus the execution time of a simple algebric kernel can lead to different execution times and trade-offs according to the strategy.... % Optimizers leverages internal models of the different contributions to decide the best strategy. Those models depends on a deep understaing of the underlying archicture and empirical characterization. 

From a computational perspective, a significant part of the workload in Transformer, and Deep Neural Network in general, consists of matrix-matrix multiplication (MatMul) operations. Thus, the efficiency of these operations that depends on the underlying hardware architecture and the specific data movement strategies employed, can greatly influence the overall performance. 
% Diverse architecture hanno proposto accelleratori dedicate al calcolo GEMV e GEMM: matrix extension, Tensor cores, systolic array. Sambanova, Groq, Cerebras, Tenstorrent 
%Particolarità di tenstorrent.

In the literature, various architectures are used for LLMs execution, ranging from general-purpose devices such as CPUs and GPUs to more specialized accelerators as FPGA-based devices. Whilst GPUs currently dominate the market, emerging architectures are being explored to further enhance computational performance and reduce energy consumption. Some examples of new solutions are given by Sambanova \cite{sambanova}, Groq \cite{groq}, Cerebras \cite{cerebras} and Tenstorrent \cite{tenstorrent}. 
In particular, Tenstorrent is producing RISC-V based accelerators designed specifically to accelerate AI workloads. They product lineup covers a broad spectrum of needs, ranging from lightweight workloads to large-scale, compute-intensive applications. Their architecture is inspired by the idea of looking at AI models as graphs, developing an architecture that capitalizes on this structure by organizing the components of the graph into a grid of processors. This arrangement allows data to flow easily among various operations, maximizing the overlap between computation and communication, leading to a promising solution that warrants further exploration. 

%Contributions
In this manuscript, we propose:
(i) A characterization of the Tenstorrent Grayskull e75 accelerator in performing MatMul kernel under different configurations, both in terms of performance and power consumption, discussing the execution model and showing the obtained results. 
(ii) With respect of the execution model our characterization shows that there is a significant performance difference between the first execution of a computational kernel and subsequent executions. Our evaluation shows that in the first run the execution time is dominated by the matrix tiling and the matrix multiplication kernels compilation, accounting for the 31\% and 66\% of the total time, respectively. In contrast, subsequent runs are primarily dominated by data transfer times (62\%). 
(iii) Considering only the kernel execution time, we characterized how processor grid size, matrix dimensions, data format and numerical fidelity impact the computational efficiency. Our results highlight substantial differences in achievable performance based on different configurations.  
(iv) A comparison with other SoA architectures, namely a Intel Sapphire Processor, a NVIDIA V100 GPU and a NVIDIA A100 GPU, showing the remarkable efficiency of Tenstorrent accelerator. 

\section{Background}
Tenstorrent develops a family of accelerators, based upon the same architecture. Among them, the Grayskull e75 is the smallest card in the lineup. Its architecture consists of a grid of 96 Tensix cores, each designed to separate communication components from computational ones, enhancing efficiency. Fabricated using 12nm process, the card features eight LPDDR4 memory channels (DRAM) positioned at the top and at the bottom of the processors grid, providing a total capacity of 8 GB and a bandwidth of 102.4 GB/sec. Operating at 1 GHz, it delivers a peak performance of 55 TFLOPs for floating-point 16. Whilst the Grayskull e75 has reduced peak performance, these are scaled up in the Wormhole family where bigger grid of Tensix Cores are interconnected at board and system level. 
\begin{figure}
    \centering
    \includegraphics[width=0.7\columnwidth]{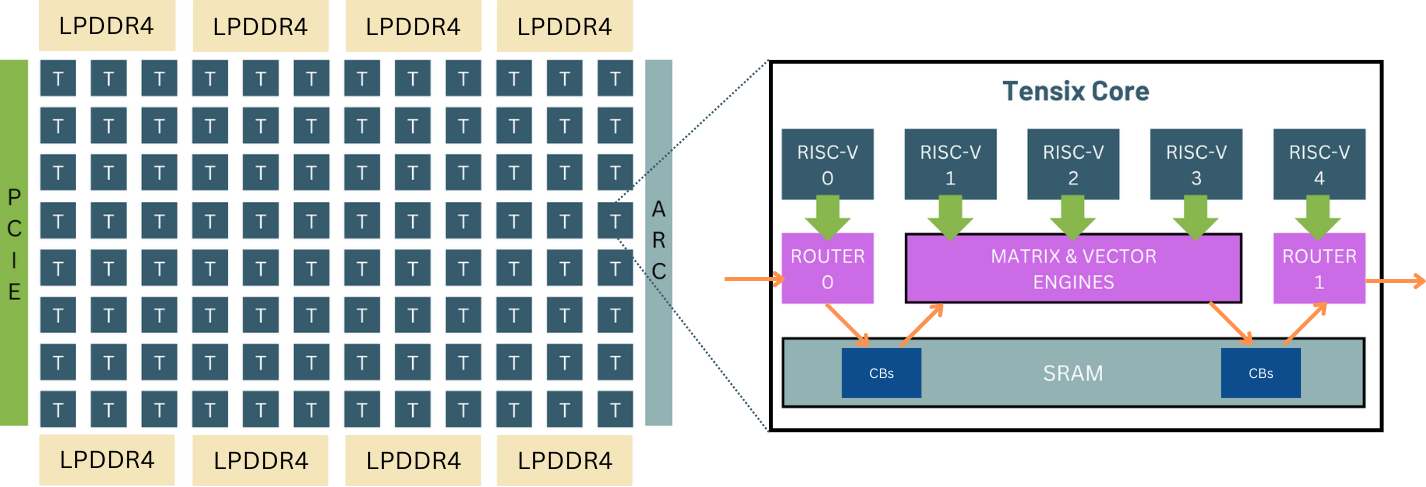}
    \caption{A schema of the Grayskull's grid architecture and of the Tensix Core.}
    \label{fig:tenstorrent_arch}
\end{figure}
In details, each Tensix Core consists of five programmable \textit{"baby"} RISC-V cores, one local SRAM memory of 1 MB (also referred as L1 memory), a SIMD Matrix \& Vector engine and the Network on Chip (NoC) routers, as shown in \autoref{fig:tenstorrent_arch}. The first and the last cores (RISC-V 0 and RISC-V 4) execute Data Movement kernels, managing the asynchronous reads and writes across the other cores' SRAMs, the external DRAM banks and the local SRAM. The remaining three cores are interfacing with the Matrix \& Vector engine, executing the Compute kernels. More specifically, the three Compute cores serve distinct roles: the first one unpacks the data, the second performs the computational kernels and the third handles data packing. 
The cores within the Tensix Core communicate using Circular Buffers (CB) in the SRAM, as shown in the figure, storing data from external memories and intermediate results.

% To further improve efficiency in computation and memory usage, Grayskull leverages several advanced techniques. One of these is the support to block sparsity of activations and model parameters, which reduces memory footprint and computation requirements. The architecture also supports dynamic precision, allowing users to set data formats and the precision of mathematical operations, called math fidelity. It is possible to set the compiler to automatically or manually adjust data formats and assign levels of mathematical fidelity to operations, reducing precision when possible.

Tenstorrent's Grayskull supports a broad range of Data Formats. Along with standard floating point formats, it supports Brain Floating Point (BF) and Block Floating Point (BFP). BFP allows the storage of blocks of 16 elements, by grouping them within a block under a shared common exponent, reducing memory footprint and improving performance. The BFP format is supported for 4, 8 and 16-bit configurations.  

Additionally, it supports four levels of Math Fidelity, which determine the computing precision by specifying the number of bits which are used for the computing operation. These levels ranges from the lower fidelity, Low Fidelity, which consumes only the most significant bits (MSB) of the mantissas of both operands, to the highest, High Fidelity 4, which consumes all the bits of both operands. The two intermediate levels, High Fidelity 2 and 3, respectively consume the MSB of the first input and the LSB of the second one, and vice-versa. Higher fidelity levels require more bits for computation, leading to an increased number of cycles per operation, resulting in lower FLOPs.

\section{Related Works}
Whilst several works have focused on characterizing the performance of SoA CPU and GPU architectures, only a few have analyzed the efficiency of Tenstorrent architecture and compared it with current SoA technologies for tensor algebra acceleration.

In \cite{brown2024acceleratingstencilstenstorrentgrayskull}, the authors explore an implementation of the Jacobi iterative method for solving Laplace's equation on the Grayskull e150, comparing the results with an Intel Xeon Platinum CPU. Their work focuses on the optimization, exploring and comparing optimal data access strategies leveraging the low-level C++ kernels. They achieved performance comparable to the Intel CPU, whilst using around five times less energy, despite the CPU running in FP32 whereas the Grayskull in BF16. 
In \cite{thüning2024attentionsramtenstorrentgrayskull} it is presented an optimization of the Attention layer using a fused kernel, achieving a significant speedup by leveraging the SRAM cores' memory. 
Additionally, Tenstorrent published a report\footnote{Available in Tenstorrent's \href{https://github.com/tenstorrent/tt-metal/blob/main/tech_reports/GEMM_FLOPS/GEMM_FLOPS.md}{GitHub repository}} in its documentation, showcasing the absolute performance of the MatMul kernel on their device Wormhole. 
\break 
Despite these initial characterizations, %which show (a) promising performance at reduce numerical precision w.r.t classical CPUs and (b) the benefit of using local memory and kernel optimizations for specific AI kernels,% 
there is a lack of a comparative analysis of simple dense tensor algebra kernel with SoA architectures featuring tensorial acceleration. %In this manuscript, we focus on evaluating the performance of Matrix <
%Talk about the tensorrent report 
In this manuscript, we provide an analysis of the MatMul kernel running on the Tenstorrent Grayskull accelerator, benchmarking its performance and comparing its efficiency against other SoA devices in terms of both performance and energy efficiency.

\section{Methodology}
In this section, we describe the Tenstorrent software stack and the software abstraction of the Tensix Cores, showing some code examples.

\subsection{TTNN}
% pytorch-like syntax
%insert figure
The TTNN Python library is a Tenstorrent's open-source library, built upon the TT-Metal software stack. It provides an API similar to PyTorch, implementing many PyTorch operations in a functional style. Under the hood, TTNN leverages C++ kernels to execute operations whilst exposing a user-friendly API.
% execution model
% insert schema 
The execution model of TTNN consists of five steps:
\begin{enumerate}
    \item Initialization of the program:
    \begin{itemize}
        \item Buffers are allocated in the L1 (SRAM) memory of each core to facilitate data synchronization.
        \item The program is compiled, generating the RISC-V binary code.
        \item The compiled program and runtime configurations (e.g., memory addresses) are loaded into L1 memory.
    \end{itemize}
    \item Creation of the DRAM buffers for data storage.
    \item Loading of the data from the host to the device's DRAM.
    \item Program execution.
    \item The results are stored back to the DRAM.
\end{enumerate}

\subsection{Software Abstraction}
Compared to PyTorch, TTNN requires users to manage Tenstorrent's memory and computation configuration. To do so, it provides control over how the data is stored across the underlying hardware, allowing users to specify the destination memory (DRAM or L1) and the tensor memory layout. 

\subsubsection{Tensor Layout}
Tensors are stored in the memory space as 2D objects by flattening the outer dimensions. For example, a tensor with dimension [1x2x4x8] is stored as [8x8]. At the lowest level, a memory block containing part of the tensor is called a page. 
There are two possible ways to map a tensor to its pages. In Row-Major layout, each row is assigned to a separate page, storing the tensor row by row from top to bottom. Alternatively, the tensor can be tiled, meaning it is stored in fixed-size blocks, the default Tenstorrent tile size is 32x32. To be used for a MatMul, the tensors must be in Tile layout.

\subsubsection{Memory Layout}
The tensor's pages can be stored in memory using two mechanisms: Interleaved or Sharded. In the Interleaved configuration, the tensor is divided into multiple pages, which are then distributed across different memory banks in a round-robin fashion. This is the default memory storage used for both the DRAM and L1 memory. Conversely, the Sharded memory configuration divides the tensor into shards and distributes them across the L1 cores' memories according to a specified mapping. This approach allows users to define a specific data distribution across the processing grid, ensuring that each core has local access to the data in its L1 memory. It is possible to define the sharding strategy (in height, width or per blocks) and orientation, which determine the order with the shards are placed on the grid.

\subsubsection{Code example}

In \autoref{lst:from_torch} is an example of how to allocate a matrix onto the device, setting its layout, the storage strategy and the format. Using this method, it is also possible to use more advanced memory strategies, as the sharded layout, showed in \autoref{lst:sharded_memory_conf}, which could be passed to the \texttt{from\_torch()} method to be applied to the input.

\lstset{basicstyle=\tiny\ttfamily, lineskip=-1pt}
\begin{lstlisting}[frame=tlrb,label=lst:from_torch,caption=Input offloading,language=Python]
    import torch 
    import ttnn
    device = ttnn.open_device(device_id=0)
    in0 = torch.randn((512,512))
    in0_t = ttnn.from_torch(
        in0,
        device=device,
        tile=ttnn.Tile((32,32)), 
        layout=ttnn.TILE_LAYOUT, 
        memory_config=ttnn.DRAM_MEMORY_CONFIG,
        dtype=ttnn.bfloat16 ) 
\end{lstlisting}

\begin{lstlisting}[frame=tlrb,label=lst:sharded_memory_conf,caption=Memory configuration definition,language=Python]
memory_config=ttnn.create_sharded_memory_config(
    (1, 1, 512, 512),
    core_grid=ttnn.CoreGrid(y=8, x=8), 
    strategy=ttnn.ShardStrategy.BLOCK,
    orientation=ttnn.ShardOrientation.ROW_MAJOR,)
\end{lstlisting}    
In \autoref{lst:kernelmatmul} is the example of a simple MatMul kernel execution, with the Math Fidelity configuration. It is possible to execute more advanced kernels by passing to the argument \texttt{program\_config} a kernel configuration.
\begin{lstlisting}[frame=tlrb,label=lst:kernelmatmul,caption=MatMul kernel execution, language=Python]
output= ttnn.matmul(
    in0_t,
    in1_t,
    dtype=ttnn.bfloat16,
    memory_config=ttnn.DRAM_MEMORY_CONFIG,
    compute_kernel_config=ttnn.GrayskullComputeKernelConfig(
        math_fidelity=ttnn.MathFidelity.HiFi4),
    #program_config=...)
\end{lstlisting}
\section{Experimental Results}
In this section, we present the characterization results for the efficiency, energy consumption and performance of the Tenstorrent Grayskull when executing the matrix-multiplication kernel. We compared these results against SoA "general-purpose" computing systems optimized for matrix multiplication and tensor linear algebra, including two NVIDIA GPUs and an Intel Xeon Platinum 8480+ processor from the Sapphire Rapids server lineup.
The tests have been conducted on the Grayskull e75, which has a peak performance of 55 TFLOPs (BF16). The Grayskull e75 is Tenstorrent's most affordable accelerator, designed for edge computing.  
%With a peak performance of 55 TFLOPs (BF16), the Grayskull e75 is Tenstorrent's most affordable accelerator, designed for edge computing. 
In Tenstorrent's lineup, other accelerators are specifically designed for large-scale computing, such as the Wormhole card, which offers a peak performance of 131 TFLOPs (BF16) and can be clustered to build high-performance computing clusters. 

\subsection{Experimental setup}
%TT connessa a xyz. The GPU xyz. CPU xyz
The Grayskull e75 is connected via PCIe 4.0 x16 to an Intel Core i7 Coffee Lake host running Ubuntu 20.04. 

To characterize the Tenstorrent accelerator, various tests have been conducted to explore the matrix multiplication execution under different conditions. We performed the following characterizations:
%The primary objective was to evaluate the benefits of offloading operations from the host to the Tenstorrent accelerator by comparing execution times. To capture and analyze various aspects, we conducted multiple families of benchmarks:
 \begin{itemize}
    \item \textbf{Offload and execution}: Analysis of the Tenstorrent execution model, consisting of the comparison between the first execution, which requires the computational kernels to be compiled, against the subsequent executions. 
    \item \textbf{Performance of the MatMul kernel}: Evaluation of the performance under different configurations, including different Data Formats, Math Fidelity and grid cores selection.
    \item \textbf{Optimized MatMul kernel}: Assessment of the performance improvement of an optimized MatMul implementation, leveraging the cores L1 memory and a more sophisticated kernel.
    \item \textbf{Energy efficiency}: Measurement and comparison of power consumption relative to performance. 
\end{itemize}
The power consumption is measured using pynvml Python module for GPUs, and TT-SMI for Grayskull, a Tenstorrent telemetry tool. Both tools report instantaneous power usage, which has been averaged over the duration of the computation.

% percentuali diferenze fra fr q sr sia nella stessa che a confronto
\subsection{Offload and execution model}
\begin{figure}
    \centering
    \includegraphics[width=0.9\columnwidth]{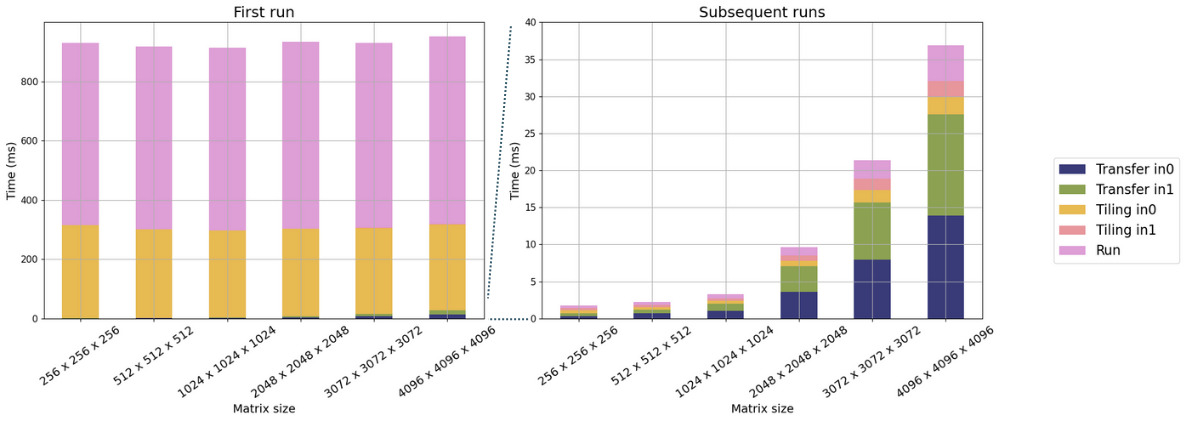}
    \caption{\footnotesize{First run and subsequent runs timings, for different matrix dimensions. The bars show the timings of the required steps: Transfer to the device of both the inputs, Tiling from Row Major to Tile layout, and the MatMul run.}}
    \label{fig:fr_vs_sb}
\end{figure}

The first execution of a kernel on the Tenstorrent Grayskull requires program compilation, which demands a substantial amount of time in comparison to the execution itself. %As shown in 
\autoref{fig:fr_vs_sb} reports the execution time of the first run and the subsequent ones.%, this is particularly evident. 
From the first run, it is possible to notice that the compilation times leads to significant overhead, which %needs to be paid in the first
happens during the first execution of the kernels: indeed we can notice that the execution time is dominated by the \textit{tiling} and the MatMul kernel \textit{run}.
%It's evident on the timing of both the two programs that need to be executed for the MatMul operation: the tiling and the MatMul kernel. 

After the first execution, the program does not need to be re-compiled, leading to a significant speedup in subsequent runs for both the \textit{tiling} and the \textit{MatMul operation} run. With respect to the tiling in the first-run plot, the tiling of the first input (\texttt{in0}) is three orders of magnitude greater than the tiling of the second matrix (\texttt{in1}), as the tiling of the second matrix will reuse the already compiled tiling program for the first one.
%, since the program needs to be compiled for the first input. 
By calculating the compilation time as the difference between the execution times of the first and subsequent executions, we obtain that the tiling kernel requires 296 ms to compile, regardless of the matrix dimensions. Whilst, the tiling execution time ranges from 0.351 ms for the 256x256 matrix to 2.256 ms for a 4096x4096 matrix. 
On the other hand, the MatMul kernel requires 620 ms for compilation, with executing times ranging from 0.328 ms to 4.783 ms for the same matrix sizes. 

In subsequent runs, data transfer timings from the host to the device become the primary overhead, accounting, on average, for 62\% of the timings. However, as the compilation, this overhead is only incurred when the matrices are located in the host. Once loaded onto the device, the operation can be executed without any transfer cost. 
Therefore, in the next experiments, we will consider only the execution time of the MatMul kernel, assuming data stationarity.

\subsection{Performance of the MatMul kernel}
\begin{table}
    \tiny
    \centering
    \renewcommand{\arraystretch}{1.0} % Increases row height
    \setlength{\tabcolsep}{12pt} % Increases column spacing
    \begin{tabular}{ l | c | c }
        \hline
        \textbf{Configuration Name} & \textbf{Data Type} & \textbf{Math Fidelity} \\
        \hline
        FP32 M4 & floating point 32 & High Fidelity 4 \\
        BF16 M4 & brain floating point 16 & High Fidelity 4 \\
        BF16 M2 & brain floating point 16 & High Fidelity 2 \\
        BFP8 M2 & Block Float 8 & High Fidelity 2 \\
        BFP8 M0 & Block Float 8 & Low Fidelity \\
        BFP4 M0 & Block Float 4 & Low Fidelity \\
    \end{tabular}
    \caption{The different configurations tested.}
    \label{configuration_table}
\end{table}
\begin{figure}
    \centering
    \begin{subfigure}{0.49\linewidth}
        \includegraphics[width=\linewidth]{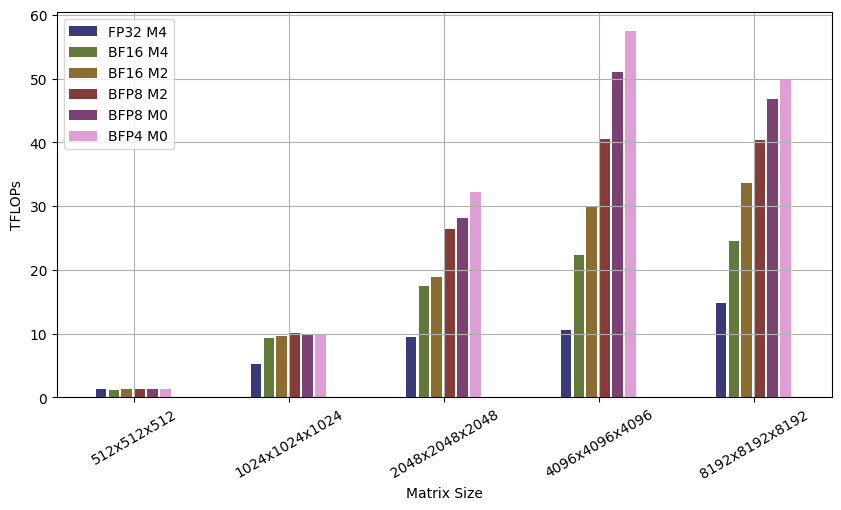}
        \caption{TFLOPs obtained using different configurations, reported in \autoref{configuration_table}}
        \label{fig:conf_comp}
    \end{subfigure}
    \hfill
    \begin{subfigure}{0.49\linewidth}
        \centering    
        \includegraphics[width=\columnwidth]{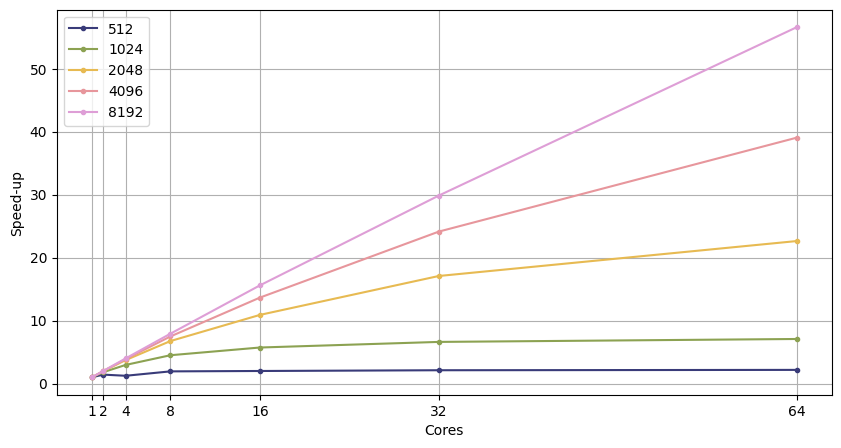}
        \caption{The MatMul speedup with respect to the single core using different set of cores, using the configuration BF16 M4.}
        \label{fig:speed_up_cores}
    \end{subfigure}
    \caption{Performance using different configurations and grid size.}
\end{figure}
We tested and compared a set of configurations reported in \autoref{configuration_table}, combining different Data Type and Math Fidelity.
In \autoref{fig:conf_comp}, we compare the MatMul TFLOPs across different configurations. As expected, performance decreases with longer Data Format and higher Math Fidelity. This is particularly evident for larger matrices dimensions, where performance ranges from 14.72 TFLOPs using floating point 32 with the highest math fidelity to 49.78 TFLOPs using Block Floating Point 4 with Low Fidelity. Considering the theoretical peak of 55 TFLOPs using BF16, the current results remain far from optimal efficiency. In the following section, we will explore optimizations of the MatMul kernel leveraging the internal L1 SRAM and the vendor-optimized MatMul kernel.

\autoref{fig:speed_up_cores} shows the speedup of the MatMul kernel for different sets of cores, ranging from 1 core to 64 cores, corresponding respectively to the grid core selection $(0, 0)$ and $(8, 8)$. 

%The full 96 cores are not reported because the selection of the full grid size $(8, 12)$ doesn't improve the performance, showing the same results of the 72 cores, thus we considered that some of the cores in the grid are not available for the kernel execution.     

As shown in the figure, smaller matrix sizes saturate the performance with few cores, whilst with larger matrix is possible to appreciate an almost linear speedup using a larger set of cores, reaching a speedup of 56x using 64 cores.

\subsection{Optimized MatMul kernel}
In addition to the default MatMul configurations, Tenstorrent's software stack allows for further performance optimization by leveraging advanced kernels that utilize sharded memory configurations, distributing the matrix across multiple cores' local SRAM. In the previous experiments, both the input matrices were allocated in DRAM using the interleaved memory storage strategy. However, performances can be further improved by sharding one of the input matrices and distributing the shards onto the L1 memory of the computing cores.
A kernel that takes advantage of this storing strategy and intermediate data reuse is the \texttt{MatmulMultiCoreReuseMultiCast} kernel, which achieves the highest performance.
\begin{figure}
\begin{subfigure}{0.49\linewidth}
    \centering
    \includegraphics[width=\linewidth]{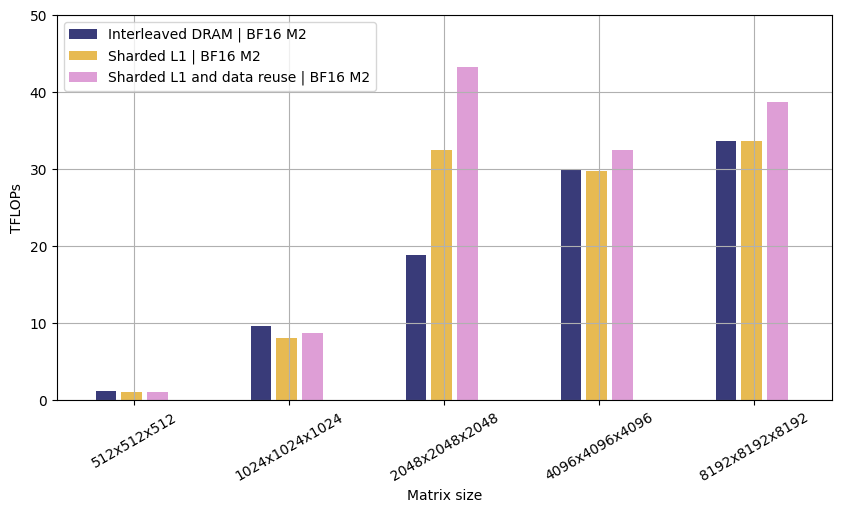}
    \caption{TFLOPs achieved with BF16 M2}
    \label{fig:opt_vs_oob}    
\end{subfigure}
\begin{subfigure}{0.49\linewidth}
    \centering
    \includegraphics[width=\linewidth]{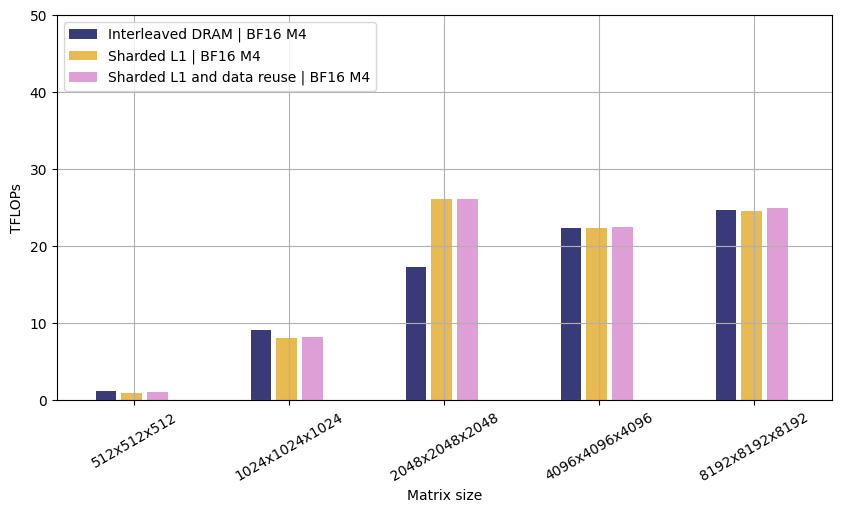}
    \caption{TFLOPs achieved with BF16 M4}
    \label{fig:opt_vs_oob_m4}    
\end{subfigure}
\caption{Comparison of TFLOPs achieved using the default kernel configuration, a sharded memory configuration and the optimized kernel with BF16 with High Fidelity 2 and 4.}
\end{figure}

In \autoref{fig:opt_vs_oob}, is shown a comparison of the default MatMul kernel (with both inputs stored in the DRAM using the interleaved configuration, shown in blue), against configurations with one input sharded in L1 memory (yellow) and the sharded memory combined with the optimized kernel. 
For smaller matrix sizes, both the optimized kernel and the L1 memory configuration exhibit similar or slightly lower performance due to the sharded memory overhead. The advantages of the memory configuration are particularly evident for the BF16 M2 configuration and the 2048x2048 MatMul, as this is the largest matrix dimension of the matrices set which can be stored in the L1 cores' memory. From the Figure, we can see that for larger matrices sizes that exceed L1 memory capacity, the shared memory optimization is no more effective, but there is still a marginal improvement in performance for the optimized kernel.

From \autoref{fig:opt_vs_oob_m4}, we can see that using the BF16 M4 configuration, the sharded memory optimization provides a similar performance improvement as observed in the M2 configuration. However, the optimized kernel does not appear particularly effective in this case.

Being the most performing one, we will use the optimized kernels for the following comparisons.
\subsubsection{Performance comparison}
\begin{figure}
    \centering
    \begin{subfigure}{0.49\linewidth}
        \centering
        \includegraphics[width=\linewidth]{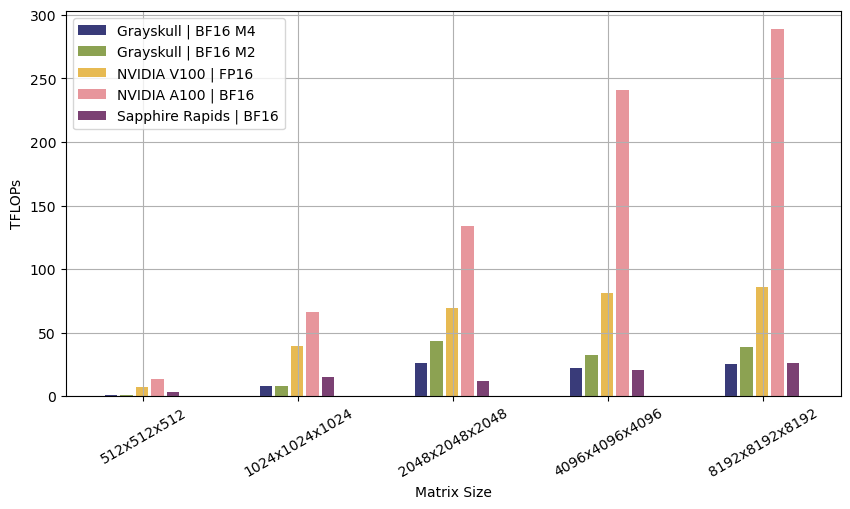}
        \caption{TFLOPS achieved in the MatMul execution.}
        \label{fig:tflops_comparison}
    \end{subfigure}
    \hfill
    \begin{subfigure}{0.49\linewidth}
        \centering
        \includegraphics[width=\linewidth]{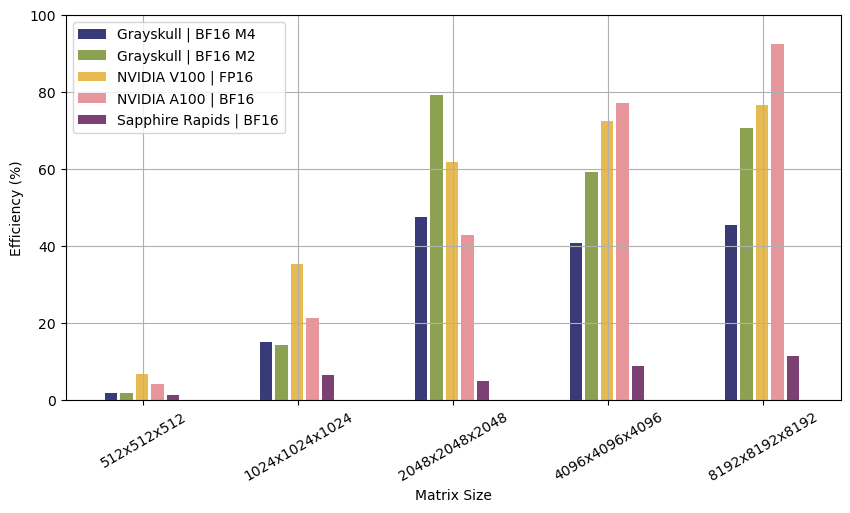}
        \caption{Efficiency calculated against peak TFLOPs.}
        \label{fig:efficiency}
    \end{subfigure}
    \caption{A comparison of the performance between the devices.}
\end{figure}

% , the A100 SXM4 40GB and the V100S PCIe 32GB, 
The obtained performance results have been compared against the PyTorch MatMul kernel execution on three SoA architectures: two GPUs, the NVIDIA A100 SXM4 40GB (w. peak BF16 throughput of 312 TFLOPs) and the NVIDIA V100S PCIe 32GB (w. peak FP16 throughput of 112 TFLOPs), and the Intel Sapphire Rapids server processor (w. peak BF16 throughput of 229 TFLOPs)\footnote{Calculated considering AMX's 1024 FLOPS/Cycle, 112 cores and 2.0 GHz frequency}.
The comparison is shown in \autoref{fig:tflops_comparison}. The Sapphire Rapids Intel processor is equipped with the Intel Advanced Matrix Extensions (AMX) to optimize MatMul execution in BF16 and INT8 formats, which are leveraged by the PyTorch's MatMul kernel. The NVIDIA GPUs obtain remarkable performance thanks to the introduction of Tensor Cores, which accelerate the MatMul computation using BF16 and FP16 formats.

As shown in the figure, the two NVIDIA GPUs dominate the performances for each matrix dimension, followed by the Grayskull accelerator and the Sapphire Rapids processor. As discussed earlier, the comparison serves as a reference to other So solutions, given that these devices are designed for different market segments. 
Nevertheless, Grayskull outperforms the Intel Sapphire Rapids processor in terms of raw performance.

In \autoref{fig:efficiency}, the efficiency is presented as the percentage of achieved TFLOPs relative to the theoretical peak TFLOPs. For smaller matrices, compute efficiency is limited but improves as matrix size increases. As shown, the Grayskull's efficiency is comparable to that of other devices, reaching a peak of 79.36\%  with 2048x2048 matrices, the largest matrix size in the test set that fits within the L1 core's memory and can benefit from sharded memory optimization. 

\subsection{Energy efficiency}
% %TODO: Explain why consumption is an important factor. Cite paper energy consumed by data movement vs computation/advantages of having a distributed architecture. Show evidence of comparable GFLOPs/Watt with GPU and superior compared to high-end CPUs but with way lower \$ x Tflops.
Energy efficiency is a critical factor for today's systems' sustainability, scalability and operation costs. Memory accesses and arithmetic operations are the most relevant terms for power and energy consumption. While energy consumed for arithmetic operations can be lowered by advance technology nodes and reduced precision numerical formats, the memory access energy consumed strongly depends on the memory technology used and distance between the memory device and the compute unit \cite{mutlu_pim}. The Tenstorrent Grayskull architecture, with its grid of Tensix Cores, stores data near the processing elements to achieve higher energy efficiency and reduce memory access time. 
%especially given the significant concern about the carbon footprint of LLMs. 
%As memory access results one of the primary factors in energy consumption, new computing architectures are being explored to enable computation closer to the data \cite{mutlu_pim}. With its grid of Tensix Cores, the Grayskull architecture stores data near to the processing elements, potentially unlocking high energy efficiency by minimizing memory access overhead. 
% \begin{figure}
%     \centering
%     \includegraphics[width=\columnwidth]{images/tflops_w_oob.png}
%     \caption{TFLOPs per Watt obtained using different configurations, reported in \autoref{configuration_table}}
%     \label{fig:tflops_watt_conf}
% \end{figure}
\begin{figure}
    \centering
    \includegraphics[width=0.5\columnwidth]{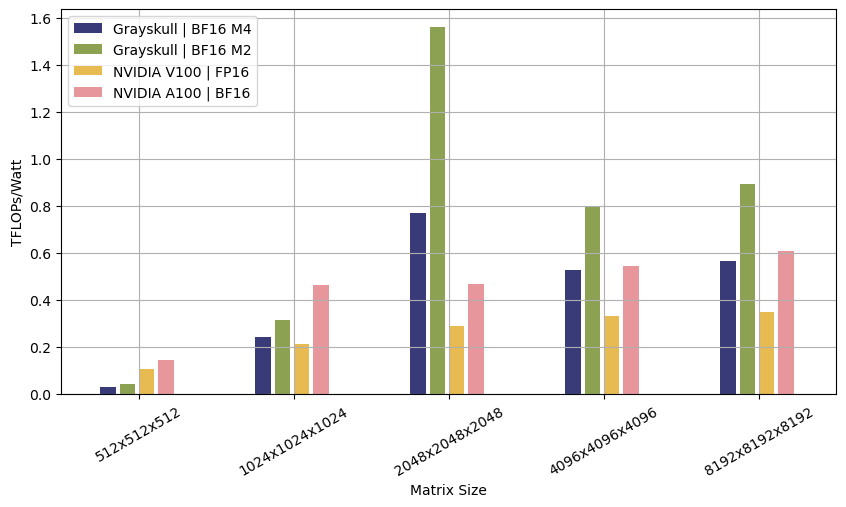}
    \caption{Comparison of achieved TFLOPs per Watt.}
    \label{fig:tflops_w_comparison}
\end{figure}

\autoref{fig:tflops_w_comparison} shows the TFLOPs per Watt achieved by the examined devices. In line with the previous considerations, the Grayskull achieves the highest efficiency, reaching a peak of 1.56 TFLOPs/Watt (BF16, M2). As previously discussed, this peak efficiency is obtained with the largest matrix size among the ones tested, which fits the grid L1 memory. %It is worth noting that at higher Mathematical Fidelity (M4) the 
It is worth noting that beyond this specific scenario, Grayskull's TFLOPs/Watt ratio remains lower than that of NVIDIA A100 when using the highest Mathematical Fidelity but surpasses it when operating at reduced precision.
\section{Conclusions}
In this manuscript, we evaluated the performance and efficiency of the Tenstorrent Grayskull e75 RISC-V accelerator in executing matrix-matrix multiplication (MatMul), a fundamental operation in deep learning. Our analysis characterized its execution model, revealing significant differences between initial and subsequent runs due to compilation and data movement overheads. We examined the impact of processor grid size, matrix dimensions, data formats and numerical fidelity on computational performance. The results demonstrate that Grayskull achieves competitive performance in terms of TFLOPs per Watt relative to SoA architectures, such as two NVIDIA GPUs (A100 and V100) and an Intel Sapphire Rapids processor. Whilst GPUs deliver higher raw throughput, Grayskull provides a promising alternative with a strong balance between performance and energy efficiency. 

Overall, this study highlights the potential of RISC-V-based accelerators in accelerating AI workloads and contributes to the ongoing discussion on efficient AI hardware design.

\begin{credits}
\subsubsection{\ackname} This activity has been supported by
the HE EU Graph-Massivizer (g.a. 101093202), DECICE (g.a.
101092582), and DARE (g.a. 101143421) projects, as well as the
Italian Research Center on High Performance Computing, Big Data,
and Quantum Computing.
\end{credits}

\bibliographystyle{splncs04}
\bibliography{ref}
\end{document}